\documentclass[journal]{IEEEtran}

\usepackage{graphicx}
\usepackage{amsmath}
\usepackage{amsfonts}
\usepackage{amssymb}
\usepackage{bbm}
\usepackage{mathrsfs}

\usepackage{epstopdf}

\newcommand{\ind}[1]{\mathbbm{1}_{\{#1\}}}   %indicator

\newcommand{\ccF}{\mathscr{F}}
\newcommand{\ccG}{\mathscr{G}}
\newcommand{\ccC}{\mathscr{C}}
\newcommand{\ccT}{\mathscr{T}}
\newcommand{\ccD}{\mathscr{D}}
\newcommand{\ccL}{\mathscr{L}}

\newcommand{\Exp}{{\sf E}}
\newcommand{\Pro}{{\sf P}}
\newcommand{\Hyp}{{\sf H}}

\newcommand{\uDi}{\underline{\Delta}_{i}}
\newcommand{\oDi}{\overline{\Delta}_{i}}
\newcommand{\uLi}{\underline{\Lambda}_{i}}
\newcommand{\oLi}{\overline{\Lambda}_{i}}

\newcommand{\cA}{\mathcal{A}}
\newcommand{\cB}{\mathcal{B}}
\newcommand{\cH}{\mathcal{H}}
\newcommand{\cN}{\mathcal{N}}
\newcommand{\cK}{\mathcal{K}}

\newcommand{\Nat}{\mathbb{N}}

\newtheorem{theorem}{Theorem}
\newtheorem{corollary}{Corollary}

\newtheorem{lemma}{Lemma}
\newtheorem{problem}{Problem}

\begin{document}

\title{Decentralized Sequential Hypothesis Testing using Asynchronous Communication\thanks{This work was supported in part by the AFOSR grant FA9550-08-1-0376.}}

\author{Georgios~Fellouris\thanks{G. Fellouris is with the Statistics Department,
Columbia University, NYC, NY, USA, e-mail: gaf106@columbia.edu.}
 and George~V.~Moustakides\thanks{G.V. Moustakides is with the Department
of Electrical and Computer Engineering, University of Patras, 26500 Rion,
Greece, e-mail: moustaki@upatras.gr.},~\IEEEmembership{Senior~Member,~IEEE}
\thanks{Manuscript received ~~~~, 2009; revised ~~~~, 2009.}
}

\markboth{IEEE Transactions on Information Theory,~Vol.~~, No.~~, ~~~2009 (submitted)}%
{Fellouris and Moustakides: Decentralized Sequential Hypothesis Testing}

\maketitle

\begin{abstract}
We present a test for the problem of decentralized sequential hypothesis testing, which is asymptotically optimum. By selecting a suitable
sampling mechanism at each sensor, communication between sensors and fusion center is asynchronous and limited to 1-bit data. The proposed SPRT-like test turns out to be order-2 asymptotically optimum in the case of  continuous time and continuous path signals, while in discrete time this strong asymptotic optimality property is preserved under proper conditions. If these conditions do not hold, then we can show optimality of order-1. Simulations corroborate the excellent performance characteristics of the test of interest.
\end{abstract}


\section{Introduction}

\IEEEPARstart{S}{equential} hypothesis testing, first introduced by Wald \cite{wald}, is one of the most classical and well-studied problems of sequential analysis with applications in areas such as industrial quality control, signal detection, design of clinical trials, etc \cite{sieg, sen}. 
In the last two decades, there has been an intense interest in the \textit{decentralized} (or \textit{distributed}) formulation of the problem \cite{tsi}-\cite{sama}. In this setup, the sequentially acquired information for decision making is distributed across a number of sensors and is transmitted to a global decision maker (fusion center), which is responsible for making the final decision. 

The main difference in the decentralized version of the problem is that the sensors are required to \textit{quantize} their observations before transmitting them to the fusion center; in other words, the sensors must send to the fusion center messages that belong to a \textit{finite alphabet} \cite{tsi}. This requirement is imposed by the need for data compression, smaller communication bandwidth and robustness of the sensor network, which are crucial issues 
in application areas such as signal processing, mobile and wireless communication, multisensor data fusion, internet security, robot networks and others \cite{ten}.  

Depending on the \textit{local memory} that the sensors possess and whether there exists \textit{feedback} from the fusion center, Veeravalli et.~al.~\cite{vbp} proposed five different configurations for the sensor network. In the same work, the authors found the optimal decentralized test -under a Bayesian setting- in the case of full feedback and local memory restricted to past decisions. 
Moreover, under a Bayesian setting, the case of no feedback and no local memory was treated in \cite{jntsi} while
the case of full local memory with no feedback in \cite{rho},\cite{venu}. However, in the last two cases no exactly optimal decentralized test has been discovered (see \cite{veer} for a review).

In this work, we  assume that the alphabet consists of two letters for all sensors, i.e. we allow the communication of only 1-bit messages. Moreover, we do not use any feedback and we consider the configuration of \textit{partial} local memory \cite{mei}. Specifically, we assume that at each time instant each sensor has access to the value of a summary statistic -that summarizes its previous observations- and uses this value, together with its current observation, in order to send a quantized signal to the fusion center. Under this configuration, an (order-1) asymptotically optimal scheme was suggested by Mei \cite{mei} under a Bayesian setting.  

Most schemes in the literature of decentralized detection require {\it synchronous} communication of the sensors with the fusion center. However, forcing distant sensors to communicate with the fusion center concurrently can be a very challenging practice. Thus, it is important to develop and analyze schemes where this communication protocol is \textit{asynchronous}. Examples of asynchronous schemes can be found in \cite{hus} and \cite{sama}.

Taking into account this consideration, we suggest that the sensors communicate with the fusion center asynchronously but also \textit{at random times}. In particular, we suggest that the times instants at which sensor
$i$ communicates with the fusion center be \textit{stopping times} that depend on the observed information at sensor $i$. We call this type of sampling \textit{adapted}. 

A special case of adapted sampling is the \textit{Lebesgue} (or \textit{level-triggered}) sampling which induces, naturally, a \textit{1-bit} communication between sensors and the fusion center. Lebesgue sampling combined with a Sequential Probability Ratio Test at the fusion center give rise to a detection structure known as Decentralized Sequential Probability Ratio Test (D-SPRT) introduced by Hussain in \cite{hus}, in a discrete time context. However, Hussain did not provide any theoretical support for this test nor evidence that it is efficient in any sense.

Our main contribution in this work consists in formulating and providing proof of asymptotic optimality of the D-SPRT, under both the discrete and the continuous time setup. Our asymptotic optimality result turns out to be stronger as compared to the scheme proposed in \cite{mei}, with simulation experiments corroborating our theoretical findings.

The case of  continuous time observations, which we analyze in Section IV, is  clearly an idealization, since in practice we cannot record the sensor observations continuously. However, studying the problem under such a setup allows us to isolate  the loss in efficiency due to discrete sampling of the underlying processes at the sensors. This provides valuable insight that leads to more efficient sampling schemes in the more realistic case of discrete time observations.

This paper is organized  as follows: Section\,I contains the Introduction. In Section\,II, we formulate the sequential hypothesis testing problem 
for the discrete and continuous time case under a centralized and decentralized setup. Moreover, we introduce the concept of adapted sampling and emphasize on Lebegsue sampling and the D-SPRT. In Section\,III we recall the main optimality results for the centralized formulation since these tests serve as a point of reference for their decentralized counterparts. Section\,IV presents the asymptotic optimality properties of D-SPRT in the context of continuous time and continuous path observations while in Section\,V we develop the same results, at the expense of a more involved analysis, for the discrete time case. In this section we also examine the notion of oversampling that ``reconciles'' the 
behavior of the discrete time D-SPRT with its continuous time version and provides some important design observations. Finally, in Section\,VI we conclude our work.

\section{Centralized versus Decentralized Sequential Testing}
\begin{figure}[b]
\centerline{\includegraphics{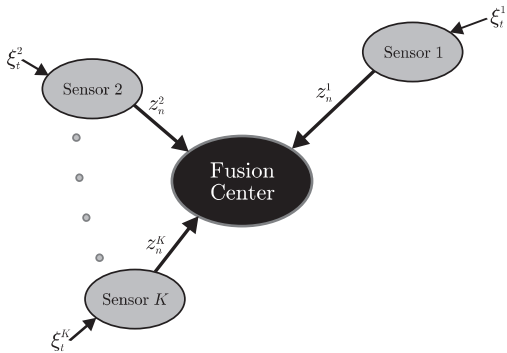}}
\caption{Schematic representation of a decentralized sensor network}
\label{fig:1}
\end{figure}
Suppose that we have a sensor network consisting of $K$ sensors as depicted in Fig.\,\ref{fig:1}.
Each sensor $i$ observes \textit{sequentially} a realization of a stochastic process $\{ \xi_{t}^i \}_{t \geq 0}$ with distribution $\Pro^{i}$. We assume that the processes $\{\xi_{t}^i\}, \ldots,\{ \xi_{t}^K\}$ are independent and we denote by $\{\ccF_{t}^i\}_{t \geq 0}$ the filtration generated by $\{\xi_t^i\}_{t \geq 0}$, where $\ccF_0^i= \{\varnothing, \Omega \}$. 
We also denote with $\Pro$ the probability measure of $\{(\xi_{t}^1, \ldots, \xi_t^K) \}_{t \geq 0}$ and by 
$\{\ccF_{t}\}_{t \geq 0}$ the filtration generated by this vector process. From the assumption of independence across sensors, we have: $\Pro= \Pro^{1} \times \ldots \times \Pro^{K}$. 

Consider now the following two hypotheses for the probability measure $\Pro$:
\begin{equation} \label{test_prob}
\Hyp_{0}: \Pro= \Pro_{0};~~~\Hyp_{1}: \Pro= \Pro_{1} , 
\end{equation}
where $\Pro_{j}= \Pro_{j}^1 \times \ldots \times \Pro_{j}^K, j=0,1,$ 
and $\Pro_j^i, \, j=0,1;~i=1, \ldots, K$ are \textit{known} probability measures. Thus $\Hyp_{0}, \Hyp_{1}$ are two \textit{simple} hypotheses.
For simplicity we also assume that each pair $\Pro_{0}^i,\Pro_{1}^i$ contains mutually absolutely continuous measures, therefore we can define the ``local'' log-likelihood ratio process at each sensor $i$ and for each time instant $t$, as follows
\begin{equation} \label{sprt_glocal}
u_t^i=\log \frac{d\Pro_{1}^i}{d\Pro_{0}^i}\left(\ccF_t^i\right);~u_{0}^i=0.
\end{equation}
Moreover, due to the independence of observations across sensors, we can write the ``global'' log-likelihood ratio $\{u_{t}\}$ in the sensor network as the sum of its local components, i.e. 
\begin{equation}
u_{t}= \log \frac{d\Pro_{1}}{d\Pro_{0}} \left(\ccF_{t}\right)= \sum_{i=1}^{K} u_t^i, \, 0 \leq t < \infty.
\label{eq:ut}
\end{equation}

Although the sensors observe sequentially the processes $\{\xi_t^i\}_{t \geq 0}$, they are allowed to communicate information to the fusion center \textit{only at a sequence of discrete times}. In particular, we assume that the fusion center receives sequentially from each sensor $i$ the data $\{z_n^i\}$ at a \textit{strictly increasing} sequence of time instants $\{\tau_n^i\}_{n \in \Nat }$. Each $\tau_n^i$ is an $\{\ccF_t^i\}$-adapted stopping time with $\tau_0^i=0$ and $\Pro_{j}(\tau_n^i < \infty) =1, \, \forall n \in \Nat , \, j=0,1$ and $i=1,\ldots,K$. We call this communication scheme \textit{adapted sampling} and we refer to the stopping times  $\{\tau_n^i\}$ as the \textit{sampling times} in sensor $i$. Each $z_n^i$ constitutes a summary of the acquired information $\ccF_{\tau_n^i}^i$ up to time $\tau_n^i$ and, as we mentioned in the Introduction, it takes values in a finite alphabet. Here we are going to assume that this set is \textit{binary}. We should also emphasize that we do not consider any feedback scheme from the fusion center towards the sensors.

Adapted sampling clearly implies \textit{asynchronous} communication between the sensors and the fusion center at \textit{random} time instants. Thus, the number of samples sent from sensor $i$ to the fusion center up to any time instant $t$ is random and in general different for each sensor. We should mention that adapted sampling is a general framework that can incorporate various sampling mechanisms already used in the literature, in particular:

\begin{itemize}

\item When $\tau_n^i- \tau_{n-1}^i= h, \, \forall n \in \Nat $, adapted sampling reduces to \textit{canonical deterministic sampling} with constant sampling period $h >0$, common to all sensors.

\item When $\{ \tau_n^i- \tau_{n-1}^i \}_{n \in \Nat }$ is a sequence of i.i.d. random variables, \textit{independent} of the observation process $\{\xi_t^i\}$, adapted sampling becomes 
\textit{independent random sampling}. For example, if  the intersampling periods $\{ \tau_n^i -\tau_{n-1}^i \}_{n \in \Nat }$ are independent and exponentially distributed with the same mean, we recover the sampling scheme suggested in \cite{sama}.

\item When the sampling times \textit{depend} on the observed sequence and are given by the following recursion
 \begin{equation} \label{lebesgue}
\tau_n^i=\inf \{t>\tau_{n-1}^i : u_{t}^i-u_{\tau_{n-1}^i}^{i} \not\in(-\uDi, \oDi) \},
\end{equation}
where $\uDi,\oDi>0$ are proper thresholds, then we call the resulting scheme \textit{Lebesgue} (or \textit{level-triggered}) sampling.
\end{itemize} 
Although not evident at first, we should emphasize that the fusion center is the recipient not only of the data sequences $\{z_n^i\}$ but also of the sampling times $\{\tau_n^i\}$ that may carry information which is relevant to the hypothesis testing problem.
Consequently, for each sensor $i$, let us define the sequence of {\it intersampling periods} $\{\delta_n^i\}_{n>0}$ where $\delta_n^i=\tau_n^i-\tau_{n-1}^i$. 

In parallel to the communication activity the fusion center, at each time instant $t$, uses all the received data up to time $t$, in order to make a decision whether to continue or stop receiving additional data. In the latter case it proceeds to make a final decision between the two hypotheses. 

Under a \textit{decentralized setup}, denote with $m_t^i$ the number of pairs $(z_n^i,\delta_n^i)$ received by the fusion center from sensor $i$ up to (and including) time $t$. We can now define the filtration $\{\ccG_t\}_{t\ge0}$ for the fusion center where $\ccG_t=\sigma\{(z_n^i,\delta_n^i), n\le m_t^i;~i=1,\ldots,K\}$ is the $\sigma$-algebra generated by all pairs $(z_n^i,\delta_n^i)$ received up to time $t$. The fusion center based on this time increasing information can use an $\{\ccG_t\}$-adapted stopping time $T$ to decide about stopping or continuing sampling. After stopping it also uses an $\ccG_{T}$-measurable decision function $d_{T}\in\{0,1\}$ to select one of the two hypotheses.

Under the \textit{centralized setup} the fusion center gains access to the {\it entire} information acquired by the sensors up to time $t$. Consequently, if $\{\ccF_t\}_{t\ge0}$ is the corresponding filtration with $\ccF_t=\sigma\{\xi_s^i,0<s\le t;~i=1,\ldots,K\}$ denoting the $\sigma$-algebra generated by all acquired information up to time $t$ then, the fusion center can use an $\{\ccF_t\}$-adapted stopping time $T$ and an $\ccF_{T}$-measurable decision function $d_T\in\{0,1\}$ to stop sampling and provide a decision between the two hypotheses.

Under both, the centralized and the decentralized formulation, our intention is to define the pair $(T,d_{T})$ optimally.
Following Wald \cite{wald}, for any $\alpha, \beta >0$, we define the class of sequential tests for which the Type-I and Type-II error probabilities are below the two levels $\alpha,\beta$ respectively, that is,
\begin{equation} \label{cab}
{\ccC}_{\alpha,\beta} = \{ (T,d_{T}):  \Pro_{0} (d_{T}=1) \leq \alpha~\mbox{and}~\Pro_{1} (d_{T}=0) \leq \beta \}.
\end{equation}
We can now define the following constrained optimization problem.
\begin{problem} 
Given $\alpha, \beta >0$ such that $\alpha + \beta <1$, find a sequential test $(\ccT,d_{\ccT}) \in {\ccC}_{\alpha,\beta}$ so that
\begin{equation} \label{wald_crit1}
\Exp_{j}[\ccT]= \inf_{(T , d_{T}) \in {\ccC}_{\alpha,\beta} } \, \Exp_{j}[T], \; \; j=0,1.
 \end{equation}
\end{problem}

If we seek the test among the $\{\ccF_t\}$-adapted schemes we refer to the optimum centralized version whereas if we limit ourselves to $\{\ccG_t\}$-adapted tests then we obtain the optimum decentralized procedure.
Note that we attempt to find a {\it single} test that \textit{simultaneously} minimizes two different criteria (the expected decision delay under the two hypotheses). It was Wald's remarkable insight that led first to conjecture \cite{wald} and then prove \cite{wolf} that a test with such extraordinary optimality property indeed exists.

Let us also introduce a second problem, proposed by Liptser and Shiryaev \cite{lip}, which constitutes a slight variant of Problem\,1.

\begin{problem} 
Given $\alpha, \beta >0$ such that $\alpha + \beta <1$, find a sequential test $(\ccT,d_{\ccT}) \in {\ccC}_{\alpha,\beta}$, so that\begin{equation} \label{wald_crit2}
\begin{split}
-\Exp_{0}[ u_{\ccT}]&=\inf_{(T , d_{T}) \in {\ccC}_{\alpha,\beta} }(-\Exp_0[ u_{T}]),\\
\Exp_{1}[ u_{\ccT}]&=\inf_{(T , d_{T}) \in {\ccC}_{\alpha,\beta} }\Exp_{1}[ u_{T}].
\end{split}
\end{equation}
\end{problem}

\noindent Recalling that $\{u_t\}$ is the running log-likelihood ratio of the two probability measures, it is clear that the two expectations $\Exp_1[u_t]$ and $-\Exp_0[u_t]$ give rise to nonnegative and increasing functions of time. These two time functions constitute, in Information Theory, a popular divergence measure known as the Kullback-Leibler (K-L) divergence. This interesting 
information theoretic criterion reduces to the usual average detection delay when the signals are i.i.d.~(in discrete time) or Brownian motions with constant drift (in continuous time).

It is clear that any decentralized scheme is bound to be inferior in performance to the optimum centralized test. This is true for two major reasons. First because a decentralized test has access to less information
($\{z_n^i\}$ being a summary of $\{\xi_t^i\}$) but also because of loss in \textit{time resolution} ($\{\tau_n^i\}$ being a sampled version of the actual time $t$). 
The main goal of our current work is to find decentralized schemes where this performance loss can be quantified and propose methods for controlling it. 

Regarding the decentralized version of Problem\,1 and 2 we must emphasize that the way it is stated, it is assumed that the sampling/quantization policy, namely the mechanism by which the pairs $\{(z_n^i,\delta_n^i)\}$ are generated from the observation sequence $\{\xi_t^i\}$, is already specified. Of course one might extend both problems by including an additional minimization over the sampling/quantization policy as well, thus optimizing all parts of the decentralized test. Finding however optimum, per se, decentralized tests that solve the extended version of the two problems turns out to be an extremely challenging task. For this reason we focus on suboptimum procedures.

To assess the quality of any decentralized test, since the optimum decentralized test is not available, we can compare it against the \textit{centralized} optimum scheme which is known in several important cases. We are in particular interested in \textit{asymptotically optimum} tests. If $\ccT$ denotes the stopping time corresponding to the optimum centralized test that solves Problem\,1 or 2 and $T$ the stopping time of a decentralized (or even centralized) competitor, then we distinguish the following degrees of asymptotic optimality\footnote{We recall the difference between the notations $\Theta(\cdot)$, $O(\cdot)$ and $o(\cdot)$. If $\omega$ is a parameter that tends to 0 or $\infty$ and $\cA(\omega),\cB(\omega)$ functions of $\omega$ then $\cA(\omega)=\Theta(\cB(\omega))$ means that $|\cA(\omega)|/|\cB(\omega)|$ is uniformly bounded away from 0 and $\infty$; $A(\omega)=O(\cB(\omega))$ that the same ratio is bounded away from $\infty$ and $A(\omega)=o(\cB(\omega))$ that $\cA(\omega)|/|\cB(\omega)|\to0$ as $\omega$ tends to 0 or $\infty$.}:

We will say that a test is \textit{asymptotically optimal of order-1}, if for $j=0,1$ and as $\alpha,\beta\to0$, we have
\begin{equation}\label{order-1}
\frac{\Exp_{j}[T]}{\Exp_{j}[\ccT]}=1+o(1),~\text{or}~
\frac{\Exp_{j}[u_{T}]}{\Exp_{j}[u_{\ccT}]}=1+o(1),
\end{equation}
for Problems\,1 and 2 respectively. 

We will say that a test is \textit{asymptotically optimal of order-2}, if for $j=0,1$ and as $\alpha,\beta\to0$, we have
\begin{equation}\label{order-2}
\Exp_{j}[T]-\Exp_{j}[\ccT]=O(1),~\text{or}~
\Exp_{j}[u_{T}]-\Exp_{j}[u_{\ccT}]=O(1),
\end{equation}
for Problems\,1 and 2 respectively. 

Finally, even though we will not consider this form of asymptotic optimality here, we define a test to be {\it asymptotically optimal of order-3}, if for $j=0,1$ and as $\alpha,\beta\to0$, we have
\begin{equation}
\Exp_{j}[T]-\Exp_{j}[\ccT]=o(1),~\text{or}~
\Exp_{j}[u_{T}]-\Exp_{j}[u_{\ccT}]=o(1).
\end{equation}
It is clear that order-3 optimality is stronger than order-2 which is stronger than order-1. Indeed order-2 implies order-1 because expected delays and K-L divergences increase without bound as $\alpha,\beta\to0$. 

In order to establish any form of asymptotic optimality, it is evident from the previous definitions that we need to recall the major results of the optimum centralized theory.

\section{Optimum Centralized Sequential Testing}

The optimization problems defined in (\ref{wald_crit1}) and \eqref{wald_crit2} are associated with the well celebrated Sequential 
Probability Ratio Test (SPRT) proposed by Wald \cite{wald}, which is defined as follows
\begin{equation} \label{sprt}
\ccT = \inf \{t>0: u_t \notin (-A,B) \} \,,\, d_{\ccT} = \left\{\begin{array}{cl} 1&\text{if}~u_{\ccT}\ge B\\0&\text{if}~u_{\ccT}\le-A,\end{array}\right.
\end{equation}
where $A,B>0$ are two thresholds and $\ccT$ is the first time the global log-likelihood ratio process $\{u_{t}\}$ leaves the open interval $(-A,B)$. The decision function $d_{\ccT}$ on the other hand is an $\ccF_{\ccT}$-measurable random variable, according to which $\Hyp_{0}$ ($\Hyp_1$) is accepted if the lower (upper) threshold is first crossed. The two thresholds $A,B$ are selected so that the two error probability constraints in (\ref{cab}) are satisfied with equality. 

In {\it continuous time} Shiryaev \cite{shiry} considered the following hypothesis testing problem
\begin{equation} \label{BMSHT} 
\Hyp_{0}: \xi_t^{i}= w_t^{i};~~~\Hyp_{1}:  \xi_t^{i}= \mu^{i} \, t +w_t^{i},
\end{equation}
where $\{w_t=(w_t^{1}, \ldots, w_t^{K})\}_{t \geq 0}$ is a $K$-dimensional Wiener process and $\mu=(\mu^{1},\ldots,\mu^K) \in \mathbf{R}^K$ are constant drifts. The local log-likelihood ratio is equal to $u_t^i=-0.5(\mu^i)^2t+\mu^i\xi_t^i$ and by summing the local components we can compute $u_t$ and apply the SPRT which is optimum in the sense of Problem\,1 and Problem\,2.

In the Brownian motion case, we have also exact formulas for the optimum performance. Specifically
\begin{equation} \label{sprt_perf}
\Exp_{0}[\ccT]= \frac{2}{\|\mu\|^2}\cH(\alpha, \beta);~~\Exp_{1}[\ccT]= \frac{2}{\|\mu\|^2}\cH(\beta, \alpha),
\end{equation}
where $\cH(x,y)= x \log( \frac{x}{1-y}) + (1-x) \log( \frac{1-x}{y})$. The two thresholds that guarantee that the two error probability constraints are satisfied with equality are given by
\begin{equation} \label{sprt_thresh}
A=\log \left( \frac{1-\alpha} {\beta} \right) \, , \,
B=\log \left( \frac{1-\beta}  {\alpha} \right).
\end{equation}

A significantly richer class of hypothesis testing problems was proposed by Liptser and Shiryaev \cite{lip} that involves It\^o processes. In particular \begin{equation} 
\Hyp_{0}: \xi_t^{i}= w_t^{i};~~~\Hyp_{1}:  \xi_t^{i}=  \int_0^t\mu_s^{i}\,ds+w_t^{i},
\label{ISHT}
\end{equation}
where, as before $\{w_{t}=(w_t^1,\ldots,w_t^K)\}_{t\ge0}$ is a $K$-dimensional Wiener process and $\{ \mu_t=(\mu_t^{1}, \ldots, \mu_t^{K}) \}_{ t \geq 0}$ is a $K$-dimensional $\{\ccF_{t}\}$-adapted process satisfying\footnote{The last condition in \eqref{assum} is known as the Novikov condition and assures that $\{e^{u_t}\}$ is a martingale. Alternative, more relaxed conditions that guarantee the martingale property can be found in \cite[Page 199]{kar}.}
\begin{equation}
\begin{split}
&\Pro_{j}\left( \int_{0}^{\infty} \|\mu_s\|^2\, ds = \infty \right) =1,\\
&\Pro_{j}\left(  \int_{0}^{t} \|\mu_s\|^2 \, ds < \infty \right)=1,\\
&\Exp_0\left[\text{exp}\left(\int_0^t\|\mu_s\|^2 \, ds\right)\right] <\infty,
\end{split}
\label{assum}
\end{equation}
for all  $t \geq 0,\,  j=0,1 $. The local log-likelihood ratio $u_t^i$ takes the form
\begin{equation}
u_t^i=-\int_0^t0.5(\mu_s^i)^2\,ds+\int_0^t\mu_s^id\xi_s^i,
\label{eq:LLRi}
\end{equation}
which again allows for the computation of $u_t$ and the application of SPRT. It is also interesting to mention that, in this particular case, the K-L divergence can be equivalently written as
\begin{equation}\label{eq:KL-Ito}
\begin{split}
-\Exp_0[u_t]&=\Exp_{0}\left[ \int_{0}^t 0.5\|\mu_s\|^2 \, ds \right]\\
\Exp_1[u_t]&=\Exp_{1}\left[ \int_{0}^t 0.5\|\mu_s\|^2 \, ds \right],
\end{split}
\end{equation}
which clearly reveals the nonnegative and time increasing nature of this alternative criterion.
As proven in \cite{lip}, under this more general setup, SPRT is optimum in the sense defined by Problem\,2
delivering the following optimal performance
\begin{equation} \label{sprt_perf_ito}
-\Exp_{0} [u_{\ccT}] = \cH(\alpha, \beta);~~
\Exp_{1} [u_{\ccT}] = \cH(\beta, \alpha),
\end{equation}
with the thresholds $A,B$ defined according to \eqref{sprt_thresh}, for the two constraints in (\ref{cab}) to be satisfied with equality. 

In {\it discrete time}, SPRT is known to be optimum in the sense of Problem\,1 and Problem\,2 when the vector sequence $\{\xi_t\}$ with $\xi_t=(\xi_t^1,\ldots,\xi_t^K)$ is i.i.d.~with independent components under both hypotheses. In particular under the two hypotheses we have
\begin{equation} \label{D-Time}
\begin{split}
\Hyp_{0}:~~& \xi_t\sim F_0(\xi^1,\ldots,\xi^K)=\prod_{i=1}^K F_0^i(\xi^i)\\
\Hyp_{1}:~~& \xi_t\sim F_1(\xi^1,\ldots,\xi^K)=\prod_{i=1}^K F_1^i(\xi^i),
\end{split}
\end{equation}
where $F_j^i(x)$ denotes the cdf of the data acquired by sensor $i$ when hypothesis $\Hyp_j$ is true and ``$\sim$'' means ``distributed according to''. For this case the local log-likelihood ratio takes the form
$$
u_t^i=\sum_{k=1}^t\log\frac{dF_1^i(\xi_k^i)}{dF_0^i(\xi_k^i)},
$$
and by summing over $i$ we can compute the global log-likelihood ratio and apply the SPRT.
The proof of optimality of SPRT was first offered by Wald and Wolfowitz in \cite{wolf}. In fact this proof constitutes the first optimality result of Sequential Analysis. We can now make the following remarks:
\begin{itemize} 

\item The SPRT has also been proven to be optimal in the case where the $\{\xi_t^i\}$ are independent homogeneous Poisson processes \cite{pes}. This problem however is not particularly interesting under the decentralized setup since an arrival at a sensor can be signaled to the fusion center using simply one bit of information.

\item In discrete time, SPRT is known to be optimum only in the i.i.d.~case. Unfortunately no analog to the It\^o class result for Problem\,2 has been developed so far.
\end{itemize}

From the optimum centralized theory we conclude that in order to apply the SPRT we need the global log-likelihood ratio $\{u_t\}$ or more precisely its local components $\{u_t^i\}$ coming from the sensors. Our goal in the next sections will be to propose efficient {\it approximations} for these processes that will replace them in the definition of SPRT thus giving rise to an SPRT-like test. The efficiency of this test will then be compared against the optimum SPRT in order to assess its asymptotic optimality.

\section{Decentralized Sequential Testing in Continuous Time}

Since we are in the continuous time case, $t$ is real taking values in $[0,\infty)$. Let us assume, but without for the moment explaining how,
that the fusion center is capable of reproducing \textit{exactly} the local log-likelihood ratio $u_{t}^i$ at the sampling instants $t=\tau_n^i$, by using only the received information $\{z_n^i\}$ from sensor $i$. It then makes sense to approximate $u_t^i$ between sampling times with its most recently reproduced value. In order to write this more formally, we recall that 
$m_t^i$ denotes the number of samples transmitted by sensor $i$ up to time $t$. Thus, at time $t$, $\tau_{m_t^i}^i$ is the most recent sampling time and $u^i_{\tau_{m_t^i}^i}$ the most recently reproduced log-likelihood value. Our suggestion is to approximate $u_t$ with $\tilde{u}_t=u^i_{\tau_{m_t^i}^i}$. We emphasize that we have exact equality between $\tilde{u}_t^i$ and $u_t^i$ at $t=\tau_n^i$, because we assume that the fusion center is capable of reproducing exactly the corresponding log-likelihood ratio at the sampling times $\{\tau_n^i\}$.

Then, the fusion center can produce an approximation $\tilde{u}_t$ for the global log-likelihood ratio $u_t$ by summing the available local approximations
\begin{equation} \label{psprt_glocal}
\tilde{u}_t = \sum_{i=1}^{K} \tilde{u}_t^i=\sum_{i=1}^K u^i_{\tau^i_{m_t^i}} \, , \; 0 \leq t < \infty,
\end{equation}
Unlike the local approximation $\tilde{u}_t^i$ which is exact at $t=\tau_n^i$, the global approximation $\tilde{u}_t$ can be exactly equal to $u_t$ at a sampling instant only if all sensors transmit synchronously, otherwise $u_t$ and $\tilde{u}_t$ will be different.

Replacing now $\{u_{t}\}$ with $\{\tilde{u}_{t}\}$ in the definition of SPRT in (\ref{sprt}), we obtain an SPRT-like test of the form
\begin{equation} \label{psprt}
\tilde{\ccT} = \inf \{t \geq 0 :\tilde{u}_t \notin (-\tilde{A},\tilde{B}) \}, ~
d_{\tilde{\ccT}} = \left\{\begin{array}{cl} 1&\text{if}~\tilde{u}_{\tilde{\ccT}}\ge \tilde{B}\\0&\text{if}~\tilde{u}_{\tilde{\ccT}}\le-\tilde{A},\end{array}\right.
\end{equation}
where again the thresholds $\tilde{A},\tilde{B}>0$ are selected to satisfy the error probability constraints with equality. The test we just described constitutes the fusion center policy we propose under the decentralized setup. Let us now explain how the fusion center can make an exact reproduction of the local log-likelihood ratios.

\subsection{Lebesgue Sampling as a Quantization Strategy}
Of course the simplest way the fusion center can reproduce the log-likelihood ratio, is by receiving the corresponding value directly from the sensor. However this would require a communication protocol that is not limited to 1-bit information. The interesting point is that, after careful consideration, the 1-bit communication constraint can be satisfied in the case of Lebesgue sampling.

Recalling that $\{\tau_n^i\}$ denotes the sequence of sampling times for sensor $i$, we have that the local log-likelihood ratio at time $\tau_n^i$ can be written as
\begin{equation}
u_{\tau_n^i}^i= \sum_{k=1}^{n} [u_{\tau_k^i}^i-u_{\tau_{k-1}^i}^i],
\label{reproduce}
\end{equation}
suggesting that the fusion center only needs the increments $u_{\tau_n^i}^i-u_{\tau_{n-1}^i}^i$ in order to recover the exact value $u_{\tau_n^i}^i$ 
at the sampling instant $\tau_n^i$. When $\{u_t^i\}$ has {\it continuous paths} and we adopt the Lebesgue sampling scheme then we observe that these 
increments can take only upon the two values $-\uDi$ or $\oDi$, since the process $u_t^i-u_{\tau_{n-1}^i}^i$ will hit at the time of sampling one of the two thresholds, due to path continuity. By assuming that the values $\uDi,\oDi$ are selected before hand and are made available to the fusion center, it then becomes easy to communicate the exact value of the increment $u_{\tau_n^i}^i-u_{\tau_{n-1}^i}^i$ by simply transmitting the following 1-bit information
\begin{equation} \label{messages}
z_n^i=\left\{
\begin{array}{cl}
1,&\mbox{if}~u_{\tau_n^i}^i-u_{\tau_{n-1}^i}^i = \oDi \\
0,&\mbox{if}~u_{\tau_n^i}^i-u_{\tau_{n-1}^i}^i = -\uDi.
\end{array}
\right.
\end{equation}
The fusion center, using the sequence $\{z_n^i\}$ and \eqref{reproduce}, can reproduce $u_t^i$ exactly at the sampling times and then form $\tilde{u}_t$ which is required in the SPRT-like test defined in \eqref{psprt}. Actually with this particular communication protocol it is possible to update directly the test statistic $\tilde{u}_t$, 
without passing through the local statistics $\tilde{u}_t^i$. Indeed the fusion center, every time it receives the 1-bit information $z_n^i$ from sensor $i$, it must simply add to the existing $\tilde{u}_t$ either $-\uDi$ or $\oDi$ depending on $z_n^i$ being 0 or 1 respectively. This observation suggests that the process $\{\tilde{u}_t\}$ is \textit{piecewise constant} exhibiting jumps every time the fusion center receives information from one or more sensors.

Lebesgue sampling in conjunction with the stopping and decision mechanism defined in \eqref{psprt} gives rise to the Decentralized Sequential Probability Ratio Test. This is in fact the continuous time version of the scheme suggested in \cite{hus} and constitutes the test that will be in the center of our attention. We emphasize that the D-SPRT is a valid \textit{decentralized} sequential test since communication is limited to 1-bit data. Before examining the optimality characteristics of the D-SPRT, let us identify certain important properties of this detection structure:

\begin{itemize}

\item Lebesgue sampling at each sensor can be seen as a local \textit{repeated} SPRT with thresholds $\uDi,\oDi$. Using \eqref{sprt_thresh} and \eqref{sprt_perf_ito} one can also prove that
\begin{equation}\label{eq:local_thres}
-\uDi=\log\frac{\Pro_1(z_n^i=0)}{\Pro_0(z_n^i=0)};~~\oDi=\log\frac{\Pro_1(z_n^i=1)}{\Pro_0(z_n^i=1)}.
\end{equation}
Consequently, for the update of the estimate $\tilde{u}_t$, the fusion center uses the log-likelihood ratio of the received bits $z_n^i$.

\item The local thresholds $\uDi,\oDi$ control the average intersampling period which is an increasing function of these two parameters. Recalling that we have two different hypotheses, we understand that the average intersampling period will depend on the true hypothesis. If we require the two 
average periods to have specific prescribed values then, using \eqref{sprt_perf} (or \eqref{sprt_perf_ito} if we want to specify the K-L divergence) and \eqref{sprt_thresh}, we can uniquely identify the local thresholds for the Brownian motion or the It\^o process case. In other data models, the two thresholds can be specified using simulations.

\item From the definition of the Lebesgue sampling scheme it is easy to see that $|u_t^i-\tilde{u}_t^i| \leq \uDi + \oDi, \, t \geq 0$, suggesting that
\begin{equation} \label{C}
|u_{t}- \tilde{u}_{t}| \leq C = \sum_{i=1}^{K} (\uDi + \oDi), t \geq 0.
\end{equation}
Thus, at any time $t$, the ``approximate'' log-likelihood ratio $\tilde{u}_{t}$ differs from the ``true'' log-likelihood ratio $u_{t}$ at most by the constant $C$.

\item As we argued above $\{\tilde{u}_t\}$ is piecewise constant. Assuming it is right continuous with left limits, the difference $\tilde{u}_t-\tilde{u}_{t\mbox{-}}$ expresses the possible jump in the process at time $t$. The largest in absolute value jump occurs when all sensors communicate at the same time and transmit data of the same sign. It is easy to verify that the maximal jump can also be bounded by
\begin{equation}
|\tilde{u}_t-\tilde{u}_{t\mbox{-}}|\le C,
\label{jump}
\end{equation}
where $C$ is defined in \eqref{C}.

\item We recall that, in addition to the data sequence $\{z_n^i\}$, each sensor transmits indirectly to the fusion center the sequence $\{\delta_n^i=\tau_n^i-\tau_{n-1}^i\}$ of intersampling periods. As we argued before, the pairs $(z_n^i,\delta_n^i)$ constitute the complete set of information received by the fusion center generating the filtration $\{\ccG_t\}$. It is also evident that the statistics of $(z_n^i,\delta_n^i)$ differ under each hypothesis suggesting that both components of the pair may carry information about the true hypothesis. We realize however that D-SPRT makes use only of the data $\{z_n^i\}$ ignoring completely the intersampling periods $\{\delta_n^i\}$. Even though this information dropout inflicts a performance loss, it turns out that it is practically advantageous. Indeed any efficient use of the pair $(z_n^i,\delta_n^i)$ would require the knowledge (or computation) of the corresponding joint pdf under the two hypotheses. Unfortunately, this is possible only for the Brownian motion model \cite{kar} and, even in this case, it is in the form of a complicated series expansion.

\end{itemize}

\subsection{Asymptotic Optimality of the D-SPRT}

Let us now establish a strong asymptotic optimality property for D-SPRT in continuous time. This is the goal of our next theorem.

\begin{theorem}\label{th:1}
Suppose that $\tilde{\ccT},d_{\tilde{\ccT}}$ is the D-SPRT test defined in \eqref{psprt}, with thresholds $\tilde{A},\tilde{B}$ selected to satisfy the error probability constraints in (\ref{cab}) with equality, then
\begin{equation} \label{eq:th1.0}
\tilde{A} \leq |\log\beta|+ C;~~\tilde{B} \leq |\log\alpha| + C.
\end{equation}
Furthermore, D-SPRT is asymptotically optimum of order-2 in the case of Problem\,1 and Problem\,2 with Brownian motion signals with constant drifts and in the case of Problem 2 with It\^o processes.
\end{theorem}

\begin{IEEEproof}
To prove \eqref{eq:th1.0}, we apply a change of measures and use \eqref{C}, this yields
\begin{equation}
\begin{split}
\beta &= \Pro_{1}(\tilde{u}_{\tilde{\ccT}} \leq -\tilde{A}) = \Exp_{0} \left[ e^{u_{\tilde{\ccT}}} \ind{\tilde{u}_{\tilde{\ccT}} \leq -\tilde{A}} \right]\\ 
&= \Exp_{0} \left[ e^{\tilde{u}_{\tilde{\ccT}}+(u_{\tilde{\ccT}}-\tilde{u}_{\tilde{\ccT}})} \ind{\tilde{u}_{\tilde{\ccT}} \leq -\tilde{A}} \right] \leq e^{-\tilde{A}+ C},
\end{split}
\end{equation}
which proves the first inequality in \eqref{eq:th1.0}. Similarly we can show the second inequality.

For order-2 optimality, we are going to prove only the case of It\^o processes and Problem\,2, since this reduces to Problem\,1 in the case of Brownian motions with constant drifts. According to the second relation in \eqref{order-2}, under hypothesis $\Hyp_0$ we need to prove that
\begin{equation}\label{eq:th1.1}
(-\Exp_0[u_{\tilde{\ccT}}])-(-\Exp_0[u_{\ccT}])=O(1).
\end{equation}
Note that the left hand side in \eqref{eq:th1.1} is always nonnegative since the SPRT, by being optimum, delivers the smallest K-L divergence. Consequently what is left to show is that the difference can be upper bounded by a constant.

Recall that $\{\tilde{u}_t\}$ is piecewise constant therefore stopping can occur only with a jump. According to \eqref{jump} the jumps of this process cannot exceed the bound $C$ defined in \eqref{C}. Before stopping, the process $\tilde{u}_t$ takes values in the interval $(-\tilde{A},\tilde{B})$ consequently, after stopping, we have
$\tilde{u}_{\tilde{\ccT}}\ge-\tilde{A}-C$. 
Using this observation, \eqref{C} and \eqref{eq:th1.0}, we can write
\begin{equation}
\begin{split}
\Exp_0[ u_{\tilde{\ccT}} ]&=\Exp_0[ \tilde{u}_{\tilde{\ccT}}+(u_{\tilde{\ccT}}-\tilde{u}_{\tilde{\ccT}})]\\
&\ge (-\tilde{A}-C)-C\ge-|\log\beta|-3C.
\end{split}
\label{eq:lastlast}
\end{equation}

From \eqref{sprt_perf_ito} we have that the performance of the SPRT, as $\alpha,\beta\to0$, satisfies $-\Exp_0[u_{\ccT}] = |\log\beta| + \alpha|\log\beta|+o(1)$. Normally $\alpha$ and $\beta$ are selected to have the same order of magnitude yielding $\alpha|\log\beta|=o(1)$, however for the validity of our theorem we can even tolerate cases where $\alpha|\log\beta|=O(1)$, that is, cases where $\alpha$ and $\beta$ are of drastically different orders of magnitudes (e.g.~$\beta=c/|\log\alpha|$). Consequently, assuming that $\alpha$ and $\beta$ converge to 0 so that $\alpha|\log\beta|+\beta|\log\alpha|=O(1)$, if in \eqref{eq:lastlast} we replace $|\log\beta|$ with the optimal SPRT performance, this proves \eqref{eq:th1.1} under $\Hyp_0$. Adopting similar arguments for the upper threshold $\tilde{B}$, we can prove \eqref{order-2} under $\Hyp_1$. This concludes the proof.
\end{IEEEproof}

\subsection{Simulation Experiments}
We now present a simulation experiment in the context of Problem\,1 with continuous time observations defined as in \eqref{BMSHT}. 
Specifically, each sensor observes a standard Wiener process under $\Hyp_0$ and a Brownian motion with a constant drift under $\Hyp_1$. We consider the case of $K=2$ sensors with the two constant drifts under $\Hyp_1$ to have the values $\mu^{1}=\mu^{2}=1$. 

\begin{figure}[!h]
\centerline{\includegraphics{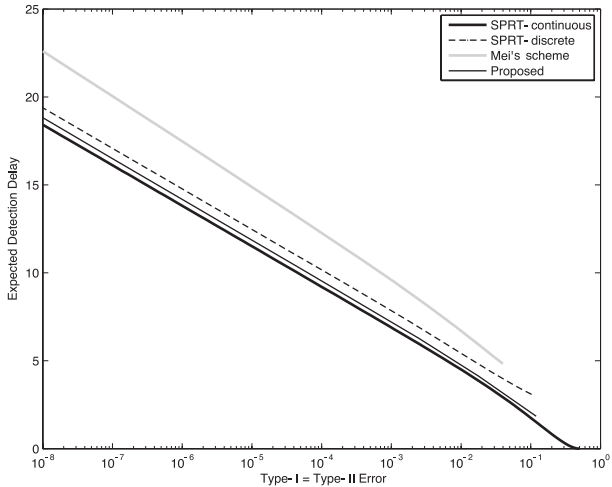}}
\caption{Relative performance of centralized and decentralized schemes in continuous time with $K=2$ sensors and testing between $\Hyp_0:$ Brownian motions with drift 0 and $\Hyp_1:$ Brownian motions with drift 1.}
\label{fig:2}
\end{figure}
We compare the D-SPRT against the continuous time (centralized) SPRT, the discrete time (centralized) SPRT and Mei's \cite{mei} decentralized test. The last two test are applied to discrete time data that are generated with canonical deterministic sampling.
For the comparison to be fair, we must equate the average intersampling periods of the Lebesgue sampling with the constant period $h$ of the canonical deterministic sampling. Selecting the local thresholds to have values $\oDi=\uDi=2$, yields $\Exp_{0}[\tau^i_{1}]=\Exp_{1}[\tau^i_{1}]=3.0464$ which must also become the value for the period of the deterministic sampling, namely $h=3.0464$.
In Fig.\,\ref{fig:2} we can see that the distance between the D-SPRT and the optimal performance remains bounded,  
which agrees with the order-2 asymptotic optimality result of Theorem\,\ref{th:1}. Mei's scheme on the other hand, known to be order-1 asymptotically optimum (see \cite{mei}), exhibits performance that slowly diverges from the optimum. 

The other important conclusion that we can draw from our graph is that the D-SPRT exhibits a distinct performance improvement over the discrete time 
SPRT which is applied after canonical deterministic sampling. We recall that this algorithm is optimum in discrete time but under the continuous time setup it is asymptotically optimum of order-2. As we argued in the Introduction, Lebesgue sampling is preferable to canonical deterministic sampling from a practical point of view since it does not require synchronization. Motivated by our simulations we can also {\it conjecture} that, even under the centralized setup, this form of sampling delivers better performance than canonical deterministic sampling.

%%%%%%%%%%%%%%%%%%%%%%%%%%%%%%%%%%%%%%%%%%%%%%%%%%%%%%%%%%%%%%%%%%%%%%%%%%%%%%%%%%%%%%%%%%%%%%%%%%%%%%%%%
%%%%%%%%%%%%%%%%%%%%%%%%%%%%%%%%%%%%%%%%%%%%%%%%%%%%%%%%%%%%%%%%%%%%%%%%%%%%%%%%%%%%%%%%%%%%%%%%%%%%%%%%%
%%%%%%%%%%%%%%%%%%%%%%%%%%%%%%%%%%%%%%%%%%%%%%%%%%%%%%%%%%%%%%%%%%%%%%%%%%%%%%%%%%%%%%%%%%%%%%%%%%%%%%%%%

\section{Decentralized Sequential Testing in Discrete Time}

We consider the same formulation as in Section\,IV only now time $t$ is discrete with $t\in\Nat$. At each sensor $i$, the process $\{\xi_t^i\}$ is i.i.d.~under the two hypotheses with corresponding cdfs $F_j^i(x),j=0,1$. Denoting with $\ell_t^i=\log(dF^i_1(\xi_t^i)/dF^i_0(\xi_t^i))$ the local log-likelihood ratio of the sample $\xi_t^i$ and assuming that $\Pro_j^i(\ell_t^i\ne1)>0$, in other words that the two densities are not equal with probability 1, we have that the global log-likelihood ratio $u_t$ is given by
\begin{equation}
u_t=\sum_{i=1}^K u_t^i=\sum_{i=1}^K\sum_{k=1}^t\ell_k^i=u_{t-1}+\sum_{i=1}^K\ell_t^i.
\end{equation}
When this definition of $u_t$ is used in \eqref{sprt}, the corresponding SPRT is optimum in the sense of Problem\,1 and 2, provided that the two thresholds $A,B$ are selected to satisfy the probability constraints in (\ref{cab})  with equality. We recall that, in discrete time, apart the i.i.d.~case, there is no other data model for which we know the solution for Problem\,2 (i.e.~there is no equivalent to the It\^o processes case).

The centralized SPRT will again become the point of reference for any decentralized test, it is therefore necessary to quantify its performance. Unfortunately in discrete time there are no exact expressions as in continuous time, we therefore need to resort to asymptotic formulas and bounds.
For the performance of SPRT we have \cite[Page 21]{sieg} the following lower bounds
\begin{equation} \label{approx_perf_sprt}
\begin{split}
-\Exp_{0}[u_{\ccT}] &\ge\cH(\alpha,\beta)=|\log\beta|+o(1),\\
\Exp_{1}[u_{\ccT}] &\ge\cH(\beta,\alpha)=|\log\alpha|+o(1),
\end{split}
\end{equation}
which replace the exact equalities of the continuous time and continuous path case depicted in \eqref{eq:KL-Ito}.

Let us now introduce a very important element in our analysis which will allow us to connect the discrete with the continuous time version presented in the previous section. 
We will assume that the ``size'' of all local log-likelihood ratios can be quantified, in an order of magnitude sense, by a finite parameter $\theta$. Normally $\theta=1$, meaning that we regard the corresponding log-likelihood ratios as being of nominal size. Here however we would like to include an additional dimension into our analysis by relating the size of the log-likelihood ratio to the error levels $\alpha,\beta$. If for example the samples $\{\xi_t^i\}$ are generated by sampling a continuous time process, then $\theta$ can be directly related to the sampling period. Our goal is to show that, for sufficiently ``small'' samples, \textit{D-SPRT enjoys the same order-2 asymptotic optimality property as its continuous time counterpart}. 
The actual size $\theta$ that can assure this interesting result, as we will show, decreases to 0, but at a \textit{much lower rate} than the two error levels $\alpha,\beta$. This suggests that with small changes in $\theta$ (coming for instance from a mild \textit{oversampling} of a continuous time process) we can obtain significant performance gains. 

It is clear that our intention is to apply the same D-SPRT scheme we introduced in the continuous time case, namely Lebesgue sampling combined with an SPRT-like test where we approximate properly the global log-likelihood ratio $u_t$. Unfortunately this transfer from the continuous to discrete time is not as straightforward as one might expect. The main reason is that with Lebesgue sampling we are no longer able to reproduce exactly the local log-likelihood ratios at the sampling times because of the \textit{overshoot effect} occurring at the local SPRT. This rather unfortunate difference is responsible for a substantial complication in the corresponding discrete time analysis.

The overshoot is of course directly related to the size of the local log-likelihood ratio of each sample. Since for our analysis the overshoot plays a very important role, it is more convenient with $\theta$ to capture the overshoot size and then, through proper conditions, to examine how $\theta$ relates to the log-likelihood ratio.

Finally, in order to avoid unnecessary complications, we will limit ourselves to the case where the two error levels $\alpha,\beta$ decrease to 0 at the same rate, meaning that the ratio $\alpha/\beta$ is uniformly bounded away from 0 and $\infty$ (or according to our definitions $\beta=\Theta(\alpha)$).

\subsection{Lebesgue Sampling and D-SPRT in Discrete Time}
In each sensor $i$, the Lebesgue sampling scheme defined in \eqref{lebesgue}, produces a sequence $\{\tau_n^i\}$ of $\{\ccF_t^i\}$-adapted stopping times, only now, due to the overshoot effect, the local SPRT statistic $u_t^i-u_{\tau_{n-1}^i}^i$ does not necessarily hit the two thresholds. Consequently the information sent over the channel can express only the \textit{side} by which the statistic $u_{\tau_n^i}^i-u_{\tau_{n-1}^i}^i$ exits the interval $(-\uDi,\oDi)$, more precisely
\begin{equation}
z_n^i=\left\{\begin{array}{cl}1&\text{if}~u_{\tau_{n}^i}^i-u_{\tau_{n-1}^i}^i\ge\oDi\\
0&\text{if}~u_{\tau_{n}^i}^i-u_{\tau_{n-1}^i}^i\le-\uDi,
\end{array}\right.
\end{equation}
which is the equivalent of \eqref{messages}. 

To this end it is only natural to ask how the fusion center should utilize the sequence $\{z_n^i\}$. In the continuous time and continuous path case, we recall that the fusion center, in view of \eqref{eq:local_thres}, uses the log-likelihood ratio of the received bits $z_n^i$ to update the estimate $\tilde{u}_t$. Consequently, in discrete time it seems only natural to use the same idea and define (as was also originally suggested in \cite{hus}) the following two quantities for each sensor 
\begin{equation}\label{Lambdas}
-\uLi=\log\frac{\Pro_1(z_n^i=0)}{\Pro_0(z_n^i=0)};~~\oLi=\log\frac{\Pro_1(z_n^i=1)}{\Pro_0(z_n^i=1)}.
\end{equation}
Both values $\uLi,\oLi$ can be precomputed either by simulations or numerically and made known to the fusion center. 

As we argued above, we are interested in the sequence of overshoots $\{\eta_n^i\}$, where
\begin{equation}\label{eq:lem01.1}
\begin{split}
\eta_n^i=&(u_{\tau_n^i}^i-u_{\tau_{n-1}^i}^i+\uDi)\ind{u_{\tau_n^i}^i-u_{\tau_{n-1}^i}^i\le-\uDi}\\
&+(u_{\tau_n^i}^i-u_{\tau_{n-1}^i}^i-\oDi)\ind{u_{\tau_n^i}^i-u_{\tau_{n-1}^i}^i\ge\oDi}.
\end{split}
\end{equation}
The maximal average overshoot size is a parameter that plays a very important role in our analysis. We define it as
\begin{equation}
\theta=\max_{j}\max_i\Exp_j[|\eta_n^i|],
\end{equation}
and we know \cite{lorden} that it is finite if $\Exp_j[(\ell_t^i)^2]<\infty,~j=0,1,~i=1,\ldots,K$.

In the continuous time and continuous path case, since there is no overshoot, the thresholds $\uDi,\oDi$ coincide with the quantities $\uLi,\oLi$. In discrete time this is no longer true. The next lemma quantifies their relative size.
\begin{lemma}\label{lem:01}
Let $\uDi,\oDi>0$ denote the thresholds for the local SPRT and $\uLi,\oLi$ be defined as in \eqref{Lambdas}, then
\begin{align}
&\uLi\ge\uDi;& &\oLi\ge\oDi \label{LgtD}\\
&\uLi=\uDi+O(\theta);& &\oLi=\oDi+O(\theta)\label{LeqDpth}
\end{align}
\end{lemma}

\begin{IEEEproof}
The proof is presented in the Appendix.
\end{IEEEproof}

The fusion center, every time it receives an information bit $z_n^i$ updates its existing statistic $\tilde{u}_t$ by either adding $-\uLi$ when $z_n^i=0$ or $\oLi$ when $z_n^i=1$. Recalling that $m_t^i$ denotes the number of bits transmitted by sensor $i$ up to time $t$, we can write for the D-SPRT statistic that $\tilde{u}_t=\sum_{i=1}^K\tilde{u}_t^i$ where
\begin{equation}\label{eq:lambda}
\tilde{u}_t^i=\sum_{n=1}^{m_t^i}\lambda_n^i;~\text{with}~\lambda_n^i=-\uLi\ind{z_n^i=0}+\oLi\ind{z_n^i=1}.
\end{equation}
The K-L information numbers of the sequence $\{\lambda_n^i\}$ play also an important role in our analysis. We have the following estimates depicted in the next lemma.
\begin{lemma}\label{lem:02}
For the K-L information numbers of the sequence $\{\lambda_n^i\}$ we can write
\begin{equation}\label{eq:lem02.1}
\begin{split}
I_0^i&=-\Exp_0[\lambda_n^i]\ge\frac{\uDi (e^{\oDi}-1)+\oDi (e^{-\uDi}-1)}{e^{\oDi}-e^{-\uDi}}>0\\
I_1^i&=\Exp_1[\lambda_n^i]\ge\frac{\uDi (e^{-\uDi}-1)+\oDi (e^{\oDi}-1)}{e^{\oDi}-e^{-\uDi}}>0.
\end{split}
\end{equation}
Additionally, if $\uDi,\oDi\to\infty$ in such a way that $\uDi/\oDi$ is bounded away from 0 and $\infty$ (i.e. $\uDi=\Theta(\oDi)$), the previous expressions simplify to
\begin{equation}\label{eq:lem02.2}
I_0^i\ge\uDi+o(1);~~I_1^i\ge\oDi+o(1).
\end{equation}
\end{lemma}

\begin{IEEEproof}
The proof is presented in the Appendix.
\end{IEEEproof}

The analysis of the classical SPRT algorithm relies on Wald's (second) identity. In order to be able to analyze the D-SPRT, it turns out that we need an equivalent result. The next lemma introduces a version of Wald's second identity that is suitable for our needs.

\begin{lemma}\label{lem:1}
Let $\{\tau_n^i\}$ denote the sequence of sampling times generated by the Lebesgue sampling scheme in sensor $i$. Consider a sequence $\{\zeta_n^i\}$ of i.i.d.~random variables where each $\zeta_n^i$ is a function of the samples $\xi_{\tau_{n-1}^i+1}^i,\ldots,\xi_{\tau_{n}^i}^i$ acquired by the sensor during the $n$th intersampling period and assume $\Exp_j[|\zeta_n^i|]<\infty$. If $T$ denotes any $\{\ccF_t\}$-adapted stopping time which is a.s.~finite with finite expectation and $m_{T}^i$ is the number of sampling times $\tau_n^i$ occurred up to and including time $T$ then, for $j=0,1$ we have
\begin{equation}\label{eq:lem1.1}
\Exp_j\left[\sum_{n=1}^{m_{T}^i+1}\zeta_n^i\right]=\Exp_j[\zeta_1^i](\Exp_j[m_{T}^i]+1).
\end{equation}
\end{lemma}

\begin{IEEEproof} The proof is presented in the Appendix.
\end{IEEEproof}

One might wonder why is it necessary to set the upper limit in \eqref{eq:lem1.1} to $m_{T}^i+1$ instead of the classical $m_{T}^i$ we encounter in Wald's original identity. Unfortunately if the upper limit is replaced by $m_{T}^i$ then in the proof (specifically in \eqref{eq:lem1.2}) the random variable $\zeta_n^i$ will be combined with $\ind{m_{T}^i\ge n}$ instead of $\ind{m_{T}^i\ge n-1}$. As it turns out, these two quantities are not necessarily independent as is the case between $\zeta_n^i$ and $\ind{m_{T}^i\ge n-1}$ and therefore Wald's identity cannot be assured. 

If we change the upper limit to $m_T^i$ then we can write two useful estimates that are an immediate consequence of Lemma\,\ref{lem:1} and are presented, without proof, in the next corollary.

\begin{corollary}\label{cor:1}
Let $\{\zeta_n^i\},T$ and $m_{T}^i$ be as in Lemma\,\ref{lem:1}, then

i). For $\zeta_n^i\ge0$ we have
\begin{equation} \label{eq:cor1.1}
\Exp_j\left[\sum_{n=1}^{m_{T}^i}\zeta_n^i\right]\le\Exp_j[\zeta_1^i](\Exp_j[m_{T}^i]+1).
\end{equation}

ii). If $\{\zeta_n^i\}$ is a sequence with $|\zeta_n^i|\le M<\infty$ for all $n$, then
\begin{equation}\label{eq:cor1.2}
\left|\Exp_j\left[\sum_{n=1}^{m_{T}^i}\zeta_n^i\right]-\Exp_j[\zeta_1^i]\Exp_j[m_{T}^i]\right|\le 2M.
\end{equation}
\end{corollary}

Unlike in continuous time, due to the overshoot effect, there is now an accumulation of errors which results in the difference $u_t-\tilde{u}_t$ being unbounded and no longer limited by a constant. However, by properly selecting the local thresholds, we will see that we can force this difference grow at a much slower pace than each of its components $u_t,\tilde{u}_t$. In turn this possibility will allow us to prove interesting asymptotic optimality properties for the D-SPRT in discrete time. Since the difference of the two statistics plays a crucial role in our analysis with the next lemma we obtain an estimate of its size.

\begin{lemma}\label{lem:2}
If $\{\eta_n^i\}$ is the sequence of overshoots generated by the the Lebesgue sampling mechanism at sensor $i$, then for any $\{\ccF_t\}$-adapted stopping time $T$ we have
\begin{equation}\label{eq:lem2.1}
\Exp_j[|u_T-\tilde{u}_T|]\le\max_i\Exp_j[|\eta_1^i|]\left(\frac{|\Exp_j[\tilde{u}_T]|+2C'}{\min_i I_j^i}+K\right)+C,
\end{equation}
where $C'=\sum_{i=1}^K(\uLi+\oLi)$ and $C=\sum_{i=1}^K(\uDi+\oDi)$. 
\end{lemma}

\begin{IEEEproof}
The proof makes use of Corollary\,1 and it is presented in the Appendix.
\end{IEEEproof}

\subsection{Asymptotic Optimality}
We have concluded the presentation of the background material that is necessary for establishing our main optimality results. 
Before going to the next theorem that introduces a key estimate for the performance of D-SPRT, we would like to introduce an additional quantity that expresses the order of magnitude of the local thresholds. We will assume that there exists a quantity $\Delta$ such that for all $i$ we have $\uDi=\Theta(\Delta)$ and $\oDi=\Theta(\Delta)$. This is necessary, because in order to establish the desired asymptotic optimality property, at some point we will require the local thresholds to tend to infinity. With this assumption all local thresholds increase at the same rate. After this clarification we can now state out next key theorem.

\begin{theorem}\label{th:2}
Let $\ccT,\tilde{\ccT}$ denote that stopping times for the centralized SPRT and D-SPRT respectively, we then have the following estimate for the thresholds of D-SPRT
\begin{equation}\label{eq:th2.1}
\tilde{A}\le|\log\beta|;~~\tilde{B}\le|\log\alpha|.
\end{equation}
Additionally, for $j=0,1$, we can write
\begin{equation}\label{eq:th2.2}
\begin{split}
|\Exp_0[u_{\tilde{\ccT}}]-\Exp_0[u_{\ccT}]|&\le\frac{\theta}{\Theta(\Delta)}|\log\beta|+\Theta(\Delta),\\
|\Exp_1[u_{\tilde{\ccT}}]-\Exp_1[u_{\ccT}]|&\le\frac{\theta}{\Theta(\Delta)}|\log\alpha|+\Theta(\Delta).
\end{split}
\end{equation}
\end{theorem}

\begin{IEEEproof}
The proof is very technical and it is presented in sufficient detail in the Appendix.
\end{IEEEproof}

We note that \eqref{eq:th2.1} is the analog of \eqref{eq:th1.0} in discrete time. In fact it constitutes a better approximation than \eqref{eq:th1.0} but at the expense of a (significantly) more involved proof. Inequality \eqref{eq:th2.2} refers to the  difference of the K-L divergences between the SPRT stopping time $\ccT$ and the D-SPRT stopping time $\tilde{\ccT}$. Since we are in the i.i.d.~case we know that the K-L divergence is proportional to the expected delay and the proportionality factor is simply the K-L information number. Theorem\,\ref{th:2} will be the starting point for establishing our asymptotic optimality results. Let us continue by first attempting to recover the continuous time analog.

\underline{\textit{Order-2 Asymptotic Optimality}}: Continuous time corresponds to ``high sampling'' or, in our terminology, to a size $\theta$ tending to 0. The question of course is what should the rate of convergence of $\theta$ towards 0 be, in order to assure the desired form of asymptotic optimality.

Assuming $\Delta=1$, in other words that the local thresholds are of the order of a nominal constant, we realize from \eqref{eq:th2.2} that we need $\theta=O(1/|\log\alpha|)=O(1/|\log\beta|)$ to reduce the right hand side in \eqref{eq:th2.2} into a quantity of the order of a constant. In other words, as we decrease the two error probabilities $\alpha,\beta$ we also need to decrease the size of the overshoot. What is however worth emphasizing is that the rate by which the size of the overshoot needs to go to 0 is \textit{much slower} than the rate of the error probabilities. This suggests that a small change in $\theta$ corresponds to a significant change in the error probabilities.

\underline{\textit{Order-1 Asymptotic Optimality}}: Of course the most crucial question is what happens if $\theta$ is considered nominal and we are allowed to play with the size $\Delta$ of the local thresholds. It is clear that in this case overly small local thresholds will induce frequent communication with the fusion center thus resulting in rapid error accumulation due to the overshoot effect. If we go through the proof of Theorem\,\ref{th:2} we realize that this part is captured by the first term in the right hand side of \eqref{eq:th2.2}. If on the other hand we use overly large local thresholds then this will generate long detection delays due to infrequent communication with the fusion center and to coarse time resolution. This part is captured by the second term in \eqref{eq:th2.2}. Clearly there is a compromising value for the local threshold size $\Delta$ that can optimize the performance of the test. 

Attempting to discover the best threshold, consider the ratio
\begin{equation} \label{eq:th2.3}
\begin{split}
0\le\frac{\Exp_j[\tilde{\ccT}]-\Exp_j[\ccT]}{\Exp_j[\ccT]}&=\frac{|\Exp_j[u_{\tilde{\ccT}}]-\Exp_j[u_{\ccT}]|}{|\Exp_j[u_{\ccT}]|}\\
&\le\frac{\theta}{\Theta(\Delta)}+\frac{\Theta(\Delta)}{|\log\alpha|}.
\end{split}
\end{equation}
If we set $\theta=1$ and let $\Delta\to\infty$ but at a rate such that $\Delta/|\log\alpha|\to0$ then the right hand side of \eqref{eq:th2.3} tends to 0 establishing order-1 asymptotic optimality. After some simple reasoning we can deduce that the best choice is $\Delta=\Theta(\sqrt{|\log\alpha|})$ which equates the two terms in \eqref{eq:th2.3}, yielding
\begin{equation}
0\le\frac{\Exp_j[\tilde{\ccT}]-\Exp_j[\ccT]}{\Exp_j[\ccT]}\le\Theta\left(\frac{1}{\sqrt{|\log\alpha}|}\right).
\end{equation}
The optimal value we obtained for $\Delta$ is the optimum local threshold size, expressed in an ``order of magnitude'' form.
Observe also that the convergence rate to 0 of the right hand side in the previous expression is of the same order as the one obtained in \cite{mei}. 

If we are now allowed to play with both, the size $\theta$ of the overshoot but also the local threshold size $\Delta$, then our previous result can be ameliorated significantly. Indeed from \eqref{eq:th2.3} we can see that the optimum size for $\Delta$ is now $\Delta=\Theta(\sqrt{\theta|\log\alpha|})$ which yields
\begin{equation}
0\le\frac{\Exp_j[\tilde{\ccT}]-\Exp_j[\ccT]}{\Exp_j[\ccT]}\le\Theta\left(\sqrt{\frac{\theta}{|\log\alpha|}}\right).
\end{equation}
Selecting $\theta$ to tend to 0 as a function of the error probability $\alpha$ makes the right hand side of the previous expression to tend to zero faster than $1/\sqrt{|\log\alpha|}$.

This theoretical result has a very useful practical implication. Specifically, we deduce that by
selecting samples which generate smaller sized overshoots, results in a D-SPRT performance improvement. As we will see in our simulations, this important characteristic is not enjoyed by Mei's decentralized scheme \cite{mei}.

\subsection{Relating the Log-likelihood ratio to the Overshoot}
Before going to our simulations let us find a way to relate the overshoot size $\Exp_j[|\eta_n^i|]$ to the log-likelihood ratio $\ell_t^i$ of a sample. Since our processes are stationary, we will consider only the case $n=1$. Recall that $\tau_0^i=u_{\tau_0^i}^i=0$, therefore
\begin{equation}
\begin{split}
\tau_1^i&=\inf\{t>0:u_t^i\notin (-\uDi,\oDi)\}\\
\eta_1^i&=(u_{\tau_1^i}^i+\uDi)\ind{u_{\tau_1^i}^i\le-\uDi}+(u_{\tau_1^i}^i-\oDi)\ind{u_{\tau_1^i}^i\ge\oDi}
\end{split}
\end{equation}
Note now that we can write $\tau_1^i=\min\{\underline{\tau}_1^i,\overline{\tau}_1^i\}$ where
\begin{equation}
\underline{\tau}_1^i=\inf\{t>0:u_t^i\le-\uDi\};~~
\overline{\tau}_1^i=\inf\{t>0:u_t^i\ge\oDi\}.
\end{equation}
Using these definitions the overshoot takes the form
\begin{equation}
\eta_1^i=(u_{\underline{\tau}_1^i}^i+\uDi)\ind{u_{\tau_1^i}^i\le-\uDi}+
(u_{\overline{\tau}_1^i}^i-\oDi)\ind{u_{\tau_1^i}^i\ge\oDi},
\end{equation}
from which we can easily deduce that
\begin{equation}
\Exp_j[|\eta_1^i|]\le\Exp_0[-(u_{\underline{\tau}_1^i}^i+\uDi)]+
\Exp_1[u_{\overline{\tau}_1^i}^i-\oDi].
\end{equation}
From \cite[Theorem\,3]{lorden} we have for $r\ge1$ that
\begin{equation}\label{eta+ell}
\begin{split}
\sup_{\uDi>0}\Exp_0[-(u_{\underline{\tau}_1^i}^i+\uDi)]&\le\left[\frac{r+2}{r+1}\frac{\Exp_0[|\ell_1^i|^{r+1}]}{|\Exp_0[\ell_1^i]|}\right]^{1/r}\\
\sup_{\oDi>0}\Exp_1[u_{\overline{\tau}_1^i}^i-\oDi]&\le\left[\frac{r+2}{r+1}\frac{\Exp_1[|\ell_1^i|^{r+1}]}{|\Exp_1[\ell_1^i]|}\right]^{1/r},
\end{split}
\end{equation}
where we have used the fact that for a nonnegative random variable $x$ and any $r\ge1$ we have $\Exp[x]\le(\Exp[x^r])^{1/r}$.

We would like to point out that \eqref{eta+ell} with $r=1$ is the most common selection for fabricating bounds for the overshoot (see \cite{sieg}). Unfortunately this value does not always produce upper bounds that tend to 0 when the corresponding log-likelihood size tends to 0. This is the reason why we had to resort to this more general form of upper bound.

\subsection{Simulation Experiments}

We illustrate our ideas by performing a simulation experiment with
$K=2$ sensors, each one observing a Brownian motion. The hypothesis testing problem we would like to solve is in the context of the problem defined in \eqref{BMSHT}, that is, under $\Hyp_0$ we have a standard Wiener process in each sensor while under $\Hyp_1$ a Brownian motion with constant drift $\mu^i$. We select the two drifts to be equal to 1, that is, $\mu^{1}=\mu^{2}=1$. 

The continuous time processes are sampled using canonical deterministic sampling with a sampling period $h$, thus generating the discrete time sequence of Normally distributed samples $\{\xi_t^i\}$ in each sensor. Clearly under $\Hyp_0$ we have that $\xi_t^i\sim\cN(0,h)$ whereas $\xi_t^i\sim\cN(h,h)$ under the alternative hypothesis $\Hyp_1$. 

The size of our samples is a function of the sampling period $h$ and tends to 0 as $h\to0$. Let us use \eqref{eta+ell} to verify that the overshoot tends to 0 as well. Forming the log-likelihood ratio we find $\ell_t^i=-0.5h+\xi_t^i$ and computing the upper bound in \eqref{eta+ell} for $r=1$ yields $3(1+0.25h)$ which, clearly, does not converge to 0 when $h\to0$. If however we select $r=2$ then the upper bound turns out to be $\Theta(h^{1/4})$ which tends to 0 with $h$. Consequently $h^{1/4}$ can play the role of $\theta$.

We compare the discrete time D-SPRT with the optimal discrete time SPRT and also with the test suggested by Mei in \cite{mei}, 
which is asymptotically optimal of order-1.
To confirm the close connection of the D-SPRT to the size of the samples (or the overshoot), we have selected two values for the sampling period, namely $h=1$ and 0.1. For the local thresholds we also considered two values, specifically $\oDi=\uDi=\Delta=1$ and 2. 

\begin{figure}
\centerline{\includegraphics{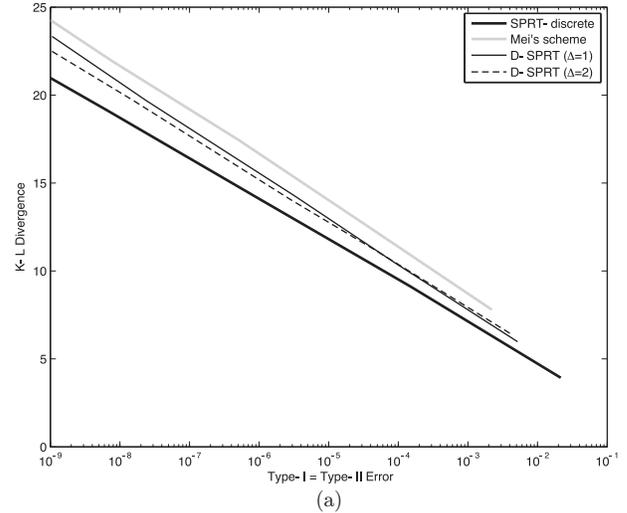}}
\vskip0.2cm
\centerline{\includegraphics{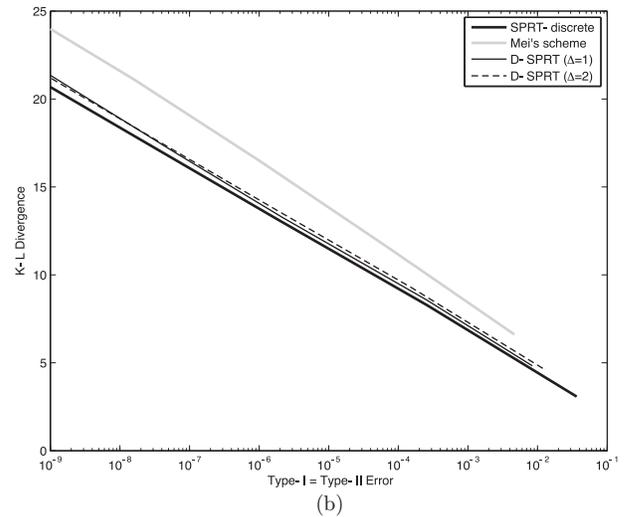}}
\caption{Relative performance of centralized and decentralized tests in discrete time with $K=2$ sensors and testing between $\Hyp_0:$ Normal $\cN(0,h)$ and $\Hyp_1:$ Normal $\cN(h,h)$ random variables with (a) $h=1.0$ and (b) $h=0.1$.}
\label{fig:3}
\end{figure}
Fig.\,\ref{fig:3} depicts the K-L divergence of the competing schemes. We recall that in this case the K-L divergence is proportional to the expected detection delay. The reason that we decided to present the former measure instead of the latter is because the K-L divergence is independent of the size of the samples while the detection delay varies drastically with this quantity (smaller samples tend to need more time to reach the same threshold).

We observe that D-SPRT exhibits a notable performance improvement when we go from the value $h=1$ to $h=0.1$. This is in complete accordance with our previous analysis since $h=0.1$ generates likelihood ratios and overshoots of smaller size than $h=1$. The optimum SPRT on the other hand and Mei's scheme are relatively insensitive to this change of size in the samples. For D-SPRT, it is basically the error accumulation expressed though the difference $|u_t-\tilde{u}_t|$ that improves as we use smaller $h$, incurring an overall performance improvement.
What is also worth emphasizing for the D-SPRT is that the communication frequency (expressed in continuous time) between the sensors and the fusion center \textit{stays relatively unchanged} under both values of $h$ while in the other two schemes it increases by a factor of 10.

Finally, in Fig.\,\ref{fig:3} we can also observe that the performance of the D-SPRT, as a function of the local threshold value $\uDi=\oDi=\Delta$, is not monotone. Indeed, case $\Delta=2$ is better than $\Delta=1$ for smaller values of $\alpha$. Additionally, the error probability values where $\Delta=2$ prevails are increasing with the size of the samples. This performance can be explained by our analysis. We recall that the optimum local threshold is $\Theta(\sqrt{\theta|\log\alpha|})$ suggesting that the error probability where any specific $\Delta$ is optimum is roughly $\alpha=\Theta(\exp(-\Delta^2/\theta))$. Consequently, a larger threshold delivers better performance at a smaller error probability and this value is an increasing function of the size $\theta$ of the samples.

%%%%%%%%%%%%%%%%%%%%%%%%%%%%%%%%%%%%%%%%%%%%%%%%%%%%%%%%%%%%%%%%%%%%%%%%%%%%%%%%%%%%%%%%%%%%%%%%%%%%%%%%%%%%%%%
%%%%%%%%%%%%%%%%%%%%%%%%%%%%%%%%%%%%%%%%%%%%%%%%%%%%%%%%%%%%%%%%%%%%%%%%%%%%%%%%%%%%%%%%%%%%%%%%%%%%%%%%%%%%%%%

\section{Conclusion}
We have presented and rigorously analyzed a decentralized scheme for sequential hypothesis testing. The detection structure relies on a local SPRT implemented at each sensor which is used for random sampling of the observed data stream. This sampling scheme naturally induces a 1-bit communication protocol between the sensors and the fusion center which is asynchronous, a very practically desirable characteristic. By performing a detailed analysis we were able to prove interesting asymptotic optimality properties for the proposed test and reveal its ability to improve performance when oversampling is used at the sensor level. Overall, our decentralized detection method exhibits performance that can be very close to the optimum centralized test, outperforming other decentralized tests of the literature.

%\section*{Acknowledgment}
%This work was supported in part by the AFOSR grant FA9550-08-1-0376.

\appendix

\textit{Proof of Lemma\,\ref{lem:01}}: To prove the lemma note that
\begin{equation}\label{eq:lem01.10}
\frac{\Pro_1(z_n^i=0)}{\Pro_0(z_n^i=0)}=\Exp_0[e^{u_{\tau_n^i}^i-u_{\tau_{n-1}^i}^i}|u_{\tau_n^i}^i-u_{\tau_{n-1}^i}^i\le-\uDi].
\end{equation}
Since
\begin{equation}
\Exp_0[e^{u_{\tau_n^i}^i-u_{\tau_{n-1}^i}^i}|u_{\tau_n^i}^i-u_{\tau_{n-1}^i}^i\le-\uDi]\le e^{-\uDi},
\end{equation}
this proves \eqref{LgtD}. For \eqref{LeqDpth}, using Jensen's inequality in \eqref{eq:lem01.10}, we can write
\begin{equation}\label{eq:lem01.11}
\Exp_0[e^{u_{\tau_n^i}^i-u_{\tau_{n-1}^i}^i}|u_{\tau_n^i}^i-u_{\tau_{n-1}^i}^i\le-\uDi]\ge
e^{-\uDi}e^{-\ccD}
\end{equation}
where
\begin{equation}
\begin{split}
\ccD&=\Exp_0[-(u_{\tau_n^i}^i-u_{\tau_{n-1}^i}^i+\uDi)|u_{\tau_n^i}^i-u_{\tau_{n-1}^i}^i\le-\uDi]\\
&=\frac{\Exp_0[-(u_{\tau_n^i}^i-u_{\tau_{n-1}^i}^i+\uDi)\ind{u_{\tau_n^i}^i-u_{\tau_{n-1}^i}^i\le-\uDi]}]}
{\Pro_0(u_{\tau_n^i}^i-u_{\tau_{n-1}^i}^i\le-\uDi)}\\
&=\frac{\Exp_0[-(u_{\tau_n^i}^i-u_{\tau_{n-1}^i}^i+\uDi)\ind{u_{\tau_n^i}^i-u_{\tau_{n-1}^i}^i\le-\uDi]}]}
{1-\Pro_0(u_{\tau_n^i}^i-u_{\tau_{n-1}^i}^i\ge\oDi)}\\
&\le\frac{\theta}{1-e^{-\oDi}},
\end{split}
\end{equation}
where in the last inequality we used the fact that the numerator is an overshoot and therefore bounded by $\theta$ and in the denominator we used Wald's approximation (which provides an upper bound) for the error probability of the local SPRT exiting from the wrong side. Replacing the bound for $\ccD$ in \eqref{eq:lem01.11}, taking the logarithm and recalling \eqref{LgtD} we conclude
\begin{equation}
0\le\uLi-\uDi\le\frac{\theta}{1-e^{-\oDi}}.
\end{equation}
Assuming that $\oDi$ is bounded away from 0, the previous right hand side becomes $O(\theta)$ and proves the lemma.
\endproof

\textit{Proof of Lemma\,\ref{lem:02}}: Let us prove the first inequality in \eqref{eq:lem02.1}. Note that
\begin{equation}
I_0^i=-\Exp_0[\lambda_n^i]=\frac{\uLi (e^{\oLi}-1)+\oLi (e^{-\uLi}-1)}{e^{\oLi}-e^{-\uLi}}>0.
\end{equation}
By direct differentiation we can verify that the function $\cK(x,y)=\{x(e^y-1)+y(e^{-x}-1)\}/(e^y-e^{-x})$ is monotonically increasing in both its arguments, when $x,y\ge0$. Consequently from \eqref{LgtD}, namely that $\uLi,\oLi$ exceed $\uDi,\oDi$ respectively, we immediately deduce the final inequality. Proving \eqref{eq:lem02.2} is straightforward.
\endproof

\textit{Proof of Lemma\,\ref{lem:1}}: For simplicity we drop the subscript $j$ that refers to the true hypothesis. We observe that
\begin{equation}\label{eq:lem1.2}
\Exp\left[\sum_{n=1}^{m_{T}^i+1}\zeta_n^i\right]=\Exp\left[\sum_{n=1}^\infty\zeta_n\ind{m_{T}^i\ge n-1}\right].
\end{equation}
Note that $\{m_{T}^i\ge n-1\}=\{T\ge\tau_{n-1}^i\}$. By recalling that $\tau_{n-1}^i$ is an $\{\ccF_t^i\}$-adapted stopping time,
this suggests that it is also $\{\ccF_t\}$-adapted. Because of the latter observation we can assess that the event $\{T\ge\tau_{n-1}^i\}$ is $\ccF_{\tau_{n-1}^i-1}$-measurable (since $\{T \ge \tau\}$ is $\ccF_{\tau-1}$-measurable, this being true even if $\tau$ is an $\{\ccF_t\}$-adapted stopping time). Consequently $\zeta_n^i$ is independent of $\ind{m_{T}^i\ge n-1}$. Interchanging summation and expectation and using independence in \eqref{eq:lem1.2}, we immediately obtain the desired equality.

The careful reader will of course argue that we cannot interchange summation and integration so freely. Indeed this is absolutely true. We can however write $\zeta_n^i=\max\{\zeta_n^i,0\}-\max\{-\zeta_n^i,0\}$ and for each component the interchange is possible requiring only $\Exp[\max\{\zeta_n^i,0\}]<\infty$ and $\Exp[\max\{-\zeta_n^i,0\}]<\infty$, which of course is satisfied iff $\Exp[|\zeta_n^i|]<\infty$, for the lemma to be true.
\endproof

\textit{Proof of Lemma\,\ref{lem:2}}: To prove \eqref{eq:lem2.1} note that $|u_t-\tilde{u}_t|\le\sum_{i=1}^K|u_t^i-\tilde{u}^i_t|$. Using \eqref{eq:lambda} we observe that we can write
\begin{equation}\label{eq:lem2.3}
|u_t^i-\tilde{u}^i_t|\le|u_t^i-u^i_{\tau_{m_t^i}}|+\sum_{n=1}^{m_t^i}|[u_{\tau_n^i}^i-u_{\tau_{n-1}^i}^i]-\lambda_n^i|.
\end{equation}
From the definition of the Lebesgue sampling we have $|u_t^i-u^i_{\tau_{m_t^i}}|\le\uDi+\oDi$. Now note that if $u_{\tau_n^i}^i-u_{\tau_{n-1}^i}^i$ exits from the lower end then $|[u_{\tau_n^i}^i-u_{\tau_{n-1}^i}^i]-\lambda_n^i|=|[u_{\tau_n^i}^i-u_{\tau_{n-1}^i}^i]+\uLi|\le|[u_{\tau_n^i}^i-u_{\tau_{n-1}^i}^i]+\uDi|$, with the last inequality coming from \eqref{LgtD}. Similarly if $u_{\tau_n^i}^i-u_{\tau_{n-1}^i}^i$ exits from the upper end then
$|[u_{\tau_n^i}^i-u_{\tau_{n-1}^i}^i]-\lambda_n^i|\le|[u_{\tau_n^i}^i-u_{\tau_{n-1}^i}^i]-\oDi|$. In both cases we see that $|[u_{\tau_n^i}^i-u_{\tau_{n-1}^i}^i]-\lambda_n^i|\le|\eta_n^i|$, with $\eta_n^i$ the overshoot defined in \eqref{eq:lem01.1}. Consequently we can further upper bound \eqref{eq:lem2.3} using the overshoot. Replacing $t$ with $T$ then taking expectation and using \eqref{eq:cor1.1} from Corollary\,\ref{cor:1}, we obtain
\begin{equation}
\begin{split}
\Exp[|u_T^i-\tilde{u}^i_T|]&\le\uDi+\oDi+\Exp[|\eta_n^i|](\Exp[m_T^i]+1)\\
&\le\uDi+\oDi+\max_i\Exp[|\eta_n^i|](\Exp[m_T^i]+1).
\end{split}
\end{equation}
Summing over $i$ yields
\begin{equation} \label{eq:lem2.10}
\Exp[|u_T-\tilde{u}_T|]\le\max_i\Exp[|\eta_n^i|]\left(\sum_{i=1}^K\Exp[m_T^i]+K\right)+C.
\end{equation}

Using now \eqref{eq:lambda} we can write
\begin{equation}
-\Exp_0[u_T^i]=-\Exp_0\left[\sum_{n=1}^{m_T^i}\lambda_n^i
\right]\ge-\Exp_0[\lambda_n^i]\Exp_0[m_T^i]-2(\uLi+\oLi),
\end{equation}
where for the last inequality we used \eqref{eq:cor1.2} of Corollary\,\ref{cor:1} and the fact that $|\lambda_n^i|\le\uLi+\oLi$. Since by definition $I_0^i=-\Exp_0[\lambda_n^i]$ is the K-L information number for the random sequence $\{\lambda_n^i\}$ we strengthen the inequality by minimizing over $i$. Summing the result over $i$ yields
\begin{equation}
-\Exp_0[u_T]\ge(\min_i I_o^i)\sum_{i=1}^K\Exp_0[m_T^i]-2C'
\end{equation}
Solving for the sum and replacing in \eqref{eq:lem2.10} yields the desired inequality under $\Hyp_0$. Similar proof applies under $\Hyp_1$.\endproof

\textit{Proof of Theorem\,\ref{th:2}}: The proof of this theorem is very challenging. In fact, as we will see, the most important part is demonstrating the validity of the estimates in \eqref{eq:th2.1}. We recall that in the synchronous case, at \textit{each} time instant $t$, we have information arriving at the fusion center from all sensors. This scenario can be easily described through i.i.d.~statistics across time. Here however, due to the asynchronous communication, this is no longer as straightforward.

In order to solve this problem, let us concentrate on one sensor (say $i$). We know that this sensor sends the sequence of bits $\{z_n^i\}$ to the fusion center but also, indirectly, the sequence of intersampling periods $\{\delta_n^i=\tau_n^i-\tau_{n-1}^i\}$. The sequence of pairs $\{(z_n^i,\delta_n^i)\}$ is adequate to fully describe sensor's $i$ transmission activity to the fusion center. Note that these pairs are i.i.d.~across time and independent across sensors.

Let us denote with $p_j^i(z,\delta)$ the joint pdf of the pair $(z_n^i,\delta_n^i)$ where, as usual, $j=0,1$ refers to the true hypothesis. We recall that $z\in\{0,1\}$ since $z_n^i$ is a 1-bit information. We can now write the joint pdf as
\begin{equation}
p_j^i(z,\delta)=\pi_j^i(0)g_j^i(\delta|0)\ind{z=0}+(1-\pi_j^i(0))g_j^i(\delta|1)\ind{z=1},
\end{equation}
where $\pi_j^i(z)=\Pro_j(z_n^i=z)$ is the probability that sensor $i$ transmits the bit $z_n^i=z$ under hypothesis $\Hyp_j$. Similarly
$g_j^i(\delta|z)$ is the pdf of $\delta_n^i$ at sensor $i$ \textit{given} that $z_n^i=z$ under hypothesis $\Hyp_j$. For example
$g_j^i(\delta|0)$ denotes the pdf of the intersampling period \textit{given} that the local SPRT exits from the lower end. The marginal pdf of the intersampling periods $\delta_n^i$ is simply 
\begin{equation}\label{eq:th2.10}
g_j^i(\delta)=\pi_j^i(0)g_j^i(\delta|0)+(1-\pi_j^i(0))g_j^i(\delta|1).
\end{equation}

Suppose now that we are at time $t$ and that the fusion center observes $m_t^i=k$ data pairs coming from sensor $i$. We have that $\{m_t^i=k\}=\{\delta_1^i+\cdots+\delta_k^i\le t<\delta_1^i+\cdots+\delta_k^i+\delta_{k+1}^i\}=\{0\le t-\tau_k^i<\delta_{k+1}^i\}$ where $\tau_k^i=\delta_1^i+\cdots+\delta_k^i$. Let us now define the likelihood of the following event: ``up to time $t$, the fusion center observes the following $m_t^i=k$ pairs $(z_1^i,\delta_1^i),\ldots,(z_k^i,\delta_k^i)$''. Using the independence of the pairs across time, we can write
\begin{equation}
\begin{split}
\Pro_j(m_t^i=k;(z_1^i,\delta_1^i),\ldots,(z_{k}^i,\delta_{k}^i))\hskip-4cm&\\
&=\Pro_j(0\le t-\tau_k^i<\delta_{k+1}^i;(z_1^i,\delta_1^i),\ldots,(z_{k}^i,\delta_{k}^i))\\
&=[1-G_j^i(t-\tau_k^i)]\left(\prod_{n=1}^kp_j^i(z_n^i,\delta_n^i)\right)\ind{\tau_k^i\le t},
\end{split}
\end{equation}
where $G_j^i(\delta)=\int_0^\delta g_j^i(x)dx$ is the cdf of $\delta_n^i$ and $g_j^i(\delta)$ is the marginal pdf defined in \eqref{eq:th2.10}. 

The previous likelihood can be decomposed as follows
\begin{equation}
\begin{split}
\Pro_j(m_t^i=k;(z_1^i,\delta_1^i),\ldots,(z_{k}^i,\delta_{k}^i))=\left(\prod_{n=1}^k\pi_j^i(z_n^i)\right)\hskip-6.5cm&\\
&\times\left([1-G_j^i(t-\tau_k^i)]\prod_{n=1}^kg_j^i(\delta_n^i|z_n^i)\ind{\tau_k^i\le t}\right).
\end{split}
\end{equation}
The first part is the likelihood of the 1-bit data $\{z_1^i,\ldots,z_k^i\}$ and the second the likelihood of the intersampling periods $\{\delta_1^i,\ldots,\delta_k^i\}$ \textit{conditioned} on the 1-bit data $\{z_1^i,\ldots,z_k^i\}$.

If $\ccG_t^i$ denotes the $\sigma$-algebra generated by the pairs $\{(z_n^i,\delta_n^i)\}$ received up to time $t$, then the likelihood ratio between the two probability measures for sensor $i$ can be written as
\begin{equation}
\begin{split}
\frac{d\Pro_1}{d\Pro_0}(\ccG_t^i)&=\left(\prod_{n=1}^{m_t^i}\frac{\pi_1^i(z_n^i)}{\pi_0^i(z_n^i)}\right)\times
\frac{g_1^i(t,\delta_1^i,\ldots,\delta_{m_t^i}^i|z_1^i,\ldots,z_{m_t^i}^i)}{g_0^i(t,\delta_1^i,\ldots,\delta_{m_t^i}^i|z_1^i,\ldots,z_{m_t^i}^i)}\\
&=e^{\tilde{u}_t^i}\times
\frac{g_1^i(t,\delta_1^i,\ldots,\delta_{m_t^i}^i|z_1^i,\ldots,z_{m_t^i}^i)}{g_0^i(t,\delta_1^i,\ldots,\delta_{m_t^i}^i|z_1^i,\ldots,z_{m_t^i}^i)},
\end{split}
\end{equation}
where
\begin{equation}
\begin{split}
g_j^i(t,\delta_1^i,\ldots,\delta_{m_t^i}^i|z_1^i,\ldots,z_{m_t^i}^i)\hskip-3cm&\\
&=[1-G_j^i(t-\tau_k^i)]\prod_{n=1}^kg_j^i(\delta_n^i|z_n^i)\ind{\tau_k^i\le t}
\end{split}
\end{equation}
expresses the likelihood of the intersampling periods conditioned on the 1-bit data, under hypothesis $\Hyp_j$. Combining all sensors and using their independence, we end up with the following likelihood ratio that refers to the complete information $\ccG_t$ received by the fusion center until time $t$
\begin{equation}
\frac{d\Pro_1}{d\Pro_0}(\ccG_t)=e^{\tilde{u}_t}\times \ccL_t
\end{equation}
with $ \ccL_t$ denoting the \textit{likelihood ratio} of the intersampling periods conditioned on the 1-bit data, namely
\begin{equation}
 \ccL_t=\prod_{i=1}^K\frac{g_1^i(t,\delta_1^i,\ldots,\delta_{m_t^i}^i|z_1^i,\ldots,z_{m_t^i}^i)}{g_0^i(t,\delta_1^i,\ldots,\delta_{m_t^i}^i|z_1^i,\ldots,z_{m_t^i}^i)}.
\end{equation}
We are now in a position to prove \eqref{eq:th2.1}. Consider the first inequality. We have
\begin{equation}
\begin{split}
\beta&=\Pro_1(d_{\tilde{\ccT}}=0)=\Exp_1[\ind{\tilde{u}_{\tilde{\ccT}}\le-\tilde{A}}]\\
&=\Exp_0[e^{\tilde{u}_{\tilde{\ccT}}}\times
 \ccL_{\tilde{\ccT}}\ind{\tilde{u}_{\tilde{\ccT}}\le-\tilde{A}}]
\le e^{-\tilde{A}}\Exp_0[ \ccL_{\tilde{\ccT}}]
=e^{-\tilde{A}}.
\end{split}
\end{equation}
The last equality is true because
\begin{equation}
\begin{split}
\Exp_0[ \ccL_{\tilde{\ccT}}]&=\Exp_0\left[\Exp_0\left[ \ccL_{\tilde{\ccT}}|z_1^1,\ldots,z_{m^1_{\tilde{\ccT}}}^1,\ldots,z_1^K,\ldots,z_{m^K_{\tilde{\ccT}}}^K\right]\right]\\
&=\Exp_0\left[\Exp_1\left[1|z_1^1,\ldots,z_{m^1_{\tilde{\ccT}}}^1,\ldots,z_1^K,\ldots,z_{m^K_{\tilde{\ccT}}}^K\right]\right]=1.
\end{split}
\end{equation}
This proves the first inequality. The second can be proven in an analogous way.

To prove the second part of the theorem, namely \eqref{eq:th2.2}, again we consider the inequality under $\Hyp_0$. Note that
\begin{equation}
\Exp_0[u_{\tilde{\ccT}}]\ge\Exp_0[\tilde{u}_{\tilde{\ccT}}]-\Exp_0[|u_{\tilde{\ccT}}-\tilde{u}_{\tilde{\ccT}}|].
\end{equation}
Using \eqref{eq:lem2.1} from Lemma\,\ref{lem:2} the inequality becomes
\begin{equation}\label{eq:th2.A2}
\Exp_0[u_{\tilde{\ccT}}]\ge(1+\Phi)\Exp_0[\tilde{u}_{\tilde{\ccT}}]-C-2\Phi C'-K\max_i\Exp_0[|\eta_n^i|],
\end{equation}
where $\Phi=(\max_i\Exp_0[|\eta_n^i|])/(\min_i I_0^i)$. 

As in the continuous time case, we have $\tilde{u}_{\tilde{\ccT}}\ge-\tilde{A}-C'$
and using \eqref{eq:th2.1} we can write $\tilde{u}_{\tilde{\ccT}}\ge-|\log\beta|-C'$ which also implies
$\Exp_0[\tilde{u}_{\tilde{\ccT}}]\ge-|\log\beta|-C'$. Replacing the latter in \eqref{eq:th2.A2} results in
\begin{equation}
\begin{split}
\Exp_0[u_{\tilde{\ccT}}]+|\log\beta|\hskip-2.1cm&\\
&\ge-\Phi|\log\beta|-(1+3\Phi)C'-C-K\max_i\Exp_0[|\eta_n^i|].
\end{split}
\end{equation}
If we replace, in the left hand side of the previous inequality, $|\log\beta|$ with the optimum performance $-\Exp_0[u_{\ccT}]$, because of \eqref{approx_perf_sprt}, we strengthen the inequality obtaining
\begin{equation}\label{eq:th2.A3}
\begin{split}
&(-\Exp_0[u_{\tilde{\ccT}}])-(-\Exp_0[u_{\ccT}])\\
&~~~\le\Phi|\log\beta|+(1+3\Phi)C'+C+K\max_i\Exp_0[|\eta_n^i|]+o(1).
\end{split}
\end{equation}

Note now that $C=\Theta(\Delta)$ and for the overshoot we have $\max_i\Exp_0[|\eta_n^i|]\le\theta$.
In our analysis we consider $\Delta$ to be, either of the order of a constant or to tend to infinity and $\theta$ to be either of the order of a constant or to tend to 0. Because of this assumption and Lemma\,\ref{lem:01} we have $\uLi,\oLi$ that are $\Theta(\Delta)$ meaning that $C'=\Theta(\Delta)$. Because of Lemma\,\ref{lem:02}, we conclude that $\min_i I_0^i\ge\Theta(\Delta)$, consequently $\Phi\le \theta/\Theta(\Delta)$. Substituting these order of magnitudes in \eqref{eq:th2.A3} yields
\begin{equation}
\begin{split}
(-\Exp_0[u_{\tilde{\ccT}}])-(-\Exp_0[u_{\ccT}])\hskip-2.5cm&\\
&=\frac{\theta}{\Theta(\Delta)}|\log\beta|+\Theta(\Delta)+O(\theta)+o(1).
\end{split}
\end{equation}
Finally due to the relative size of $\Delta$ and $\theta$ we can also conclude that $\Theta(\Delta)+O(\theta)+o(1)=\Theta(\Delta)$ which proves the desired version of the inequality. Similar steps can be applied to prove the theorem under hypothesis $\Hyp_1$.\endproof

\end{document}